# InferA: A Smart Assistant for Cosmological Ensemble Data

Justin Z. Tam
Los Alamos National Laboratory
Los Alamos, United States
justinztam@lanl.gov

Pascal Grosset
Los Alamos National Laboratory
Los Alamos, United States
pascalgrosset@lanl.gov

Divya Banesh
Los Alamos National Laboratory
Los Alamos, United States
dbanesh@lanl.gov

Nesar Ramachandra
Argonne National Laboratory
Lemont, United States
nramachandra@anl.gov

Terece L. Turton
Los Alamos National Laboratory
Los Alamos, United States
tlturton@lanl.gov

James Ahrens
Los Alamos National Laboratory
Los Alamos, United States
ahrens@lanl.gov

## ABSTRACT

Analyzing large-scale scientific datasets presents substantial challenges due to their sheer volume, structural complexity, and the need for specialized domain knowledge. Automation tools, such as PandasAI, typically require full data ingestion and lack context of the full data structure, making them impractical as intelligent data analysis assistants for datasets at the terabyte scale. To overcome these limitations, we propose InferA, a multi-agent system that leverages large language models to enable scalable and efficient scientific data analysis. At the core of the architecture is a supervisor agent that orchestrates a team of specialized agents responsible for distinct phases of the data retrieval and analysis. The system engages interactively with users to elicit their analytical intent and confirm query objectives, ensuring alignment between user goals and system actions. To demonstrate the framework's usability, we evaluate the system using ensemble runs from the HACC cosmology simulation which comprises several terabytes.

# 1 INTRODUCTION

With the advent of exascale computing, scientific simulations now have access to unprecedented levels of computational power. This capability enables the generation of data at a scale and complexity that far exceeds previous norms. One of the most data-intensive domains is cosmology, where simulations model the evolution of the universe by tracking interactions among trillions of particles over hundreds of timesteps. Simulations, such as the Hardware/Hybrid Accelerated Cosmology Code (HACC) [12, 15], produce semantically rich outputs that include not only particles but also higher-level structures such as dark-matter halos and filaments, as shown in Fig. 1. Furthermore, these simulations are often performed as ensembles, as shown in Fig. 2—multiple runs with different initial conditions, further compounding data volume and complexity. As a result, data from HACC are of the order of petabytes, requiring an impractical amount of human time for analysis.

Large language models (LLMs), such as ChatGPT [1] and Ollama [21], are now often used for analyzing data and generating analysis scripts [24, 25]. With prompt engineering, researchers can easily request Python scripts to perform statistical queries or generate plots over structured datasets. However, applying LLMs to large-scale scientific data presents unique challenges. In the case of cosmological simulations, each time step may be described by hundreds of parameters, quickly exceeding the context window limits of current LLMs. Moreover, the types of queries relevant to cosmology, such as identifying the largest dark-matter halo and tracing its evolution, are far more specialized than those typically encountered in general-purpose analytics, making automatic code synthesis difficult and ineffective without domain-specific tuning.

Another emerging approach involves tools such as PandasAI [29], which combine LLMs with tabular data libraries to support natural language querying. However, these tools generally require the full dataset to be in memory prior to analysis. This requirement renders them impractical for use with large-scale scientific simulations. For example, a single time step from the Last Journey simulation [16], available via the HACC Data Portal [4], is approximately 540 GB. With over 500 time steps in a full run, ingesting such data would i) be very time consuming, and ii) require more memory than usually available on a single compute node of a high performance computing (HPC) system.

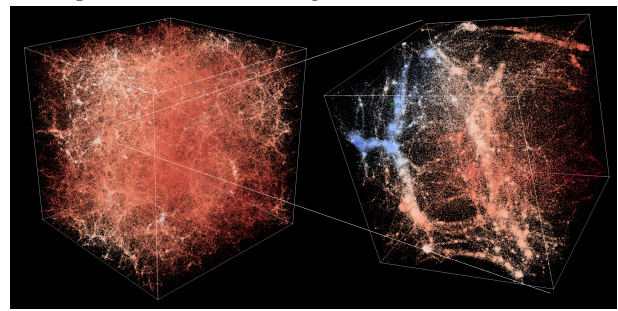

**Figure 1: A HACC simulation with 1,073,726,359 particles with a region zoomed out that shows cluster of dark matter particles (halos) and cosmic filaments.**

These limitations underscore the need for a more adaptive and scalable analysis framework tailored to high-dimensional, structured scientific data. In response, we propose a multi-agent analysis framework, **InferA** (**Infer**ence through **A**I), designed specifically to assist with large-scale ensemble data analysis of HACC data. Multi-agent systems consist of multiple autonomous agents, each responsible for a specialized task, working under the coordination of a central supervisor agent. This architecture enables modular reasoning, task delegation, and integration of domain-specific tools while maintaining global coherence and context. Our main contributions are

- A natural-language assistant capable of analyzing ensemble-scale cosmological simulation data;
- The ability to safely interact with and reason over extremely large scientific datasets without full data ingestion; and,
- Complete provenance tracking to support result verification and reproducibility by domain experts.

A fundamental requirement in scientific research is reproducibility, which is closely tied to provenance [11]. Provenance enables the interpretation and understanding of results by providing a clear record of the sequence of steps that produced them. Recent efforts in machine learning also emphasize reproducibility and explainability through transparent experimental design and result tracking [14, 26]. In line with these practices, InferA embeds fine-grained provenance tracking to support auditing, verification, and reproducibility of all analytical decisions [31]. The code for InferA is available at https://github.com/lanl/InferA.

In this paper, we target the data products of HACC simulations, and specifically the extension that resolves hydrodynamics in the large-scale structure formation of the universe, as shown in Fig. 2. HACC has a gravitational N-body solver for dark matter particles, as well as a modern smoothed-particle hydrodynamics component, to accurately model baryonic effects in cosmology simulations. To capture unresolved physics under the resolution limits, numerous sub-grid physics models are employed. We use an ensemble of HACC simulations with five varied sub-grid parameters: the stellar feedback energy fraction $f_{SN}$, the logarithm of the stellar feedback kick velocity $\log(v_{SN})$, the active galactic nuclei (AGN) feedback temperature jump $\log(T_{AGN})$, the slope $\beta_{BH}$ controlling the density-dependent boost to black hole accretion, and the AGN seed mass $M_{seed}$. Each simulation run, with 625 time-evolution snapshots, is approximately 350 GB. The dataset contains information about dark matter particles, dark matter halos, galaxies, and core particles, among other entities. These simulations, developed and run by the HACC team at Argonne National Laboratory (ANL), help in understanding the underlying nature of dark matter and galactic physics, and provide synthetic data to be compared against ground- and space-based telescope surveys.

## 2 RELATED WORK

Artificial intelligence (AI) methods are significantly enhancing the capacity to analyze scientific data. The emergence of intelligent agents powered by LLMs enables natural-language interfaces for literature exploration, code generation, and visualization. Several surveys provide broad overviews of tools and methodologies in this domain [5, 28, 30]. Lightweight tools such as PandasAI are commonly used for general-purpose data analytics, while more advanced systems such as Data Interpreter [17] and Data-Copilot [32] combine agent-based reasoning with LLMs for structured scientific data analysis. While these frameworks offer task decomposition and iterative reasoning, compared to InferA they lack the ability to process large-scale simulation datasets and do not incorporate robust provenance mechanisms.

Recent advances in AI have shown the potential of LLMs and natural language interfaces for scientific visualization. Systems such as LightVA [34] and PhenoFlow [18] enable interactive exploration of complex datasets, while modular designs like Chat2VIS [20] and FREYR [13] decompose visualization tasks into specialized components. Multi-agent systems, including VizGenie [6], NLI4VolVis [3], and ParaView-MCP [19] combine domain-specific retrieval-augmented generation (RAG) and semantic understanding for automated visualization pipelines. However, these approaches focus on visualization and do not address scalable analytics over large-scale, ensemble scientific datasets—the core goal of InferA.

LLMs are increasingly being applied in cosmology, supporting tasks such as code generation, data analysis, and scientific writing [10]. Domain-specialized models such as *AstroSage* enable natural language querying to access astronomy and cosmology-specific knowledge [9]. Similarly, SimAgents extracts simulation parameters from published literature, bridging the gap between textual descriptions and executable cosmological simulation setups [33]. The work most closely related to ours is *The AI Cosmologist*, which employs a multi-agent architecture to automate machine learning workflows and data analysis in cosmological research [22]. However, unlike InferA, which supports natural-language-driven analysis over ensemble-scale simulation data, The AI Cosmologist is primarily focused on automating predefined workflows and operates on comparatively smaller datasets.

## 3 METHODOLOGY

We implement a two-stage workflow as illustrated in Fig. 3. The first stage involves *planning*, where an initial analysis plan is generated from the user's query and iteratively refined through human feedback. This iterative feedback loop is crucial for accurately translating user intent into actionable analysis steps. Once the plan is finalized, the second stage, *analysis*, begins, in which a supervisor

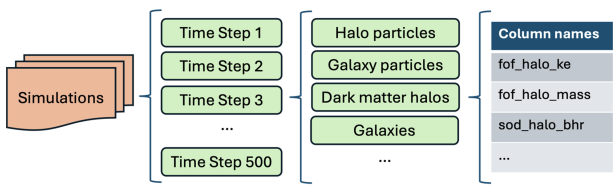

**Figure 2: Each HACC simulation has several timesteps with galaxies, halos and raw particles.**

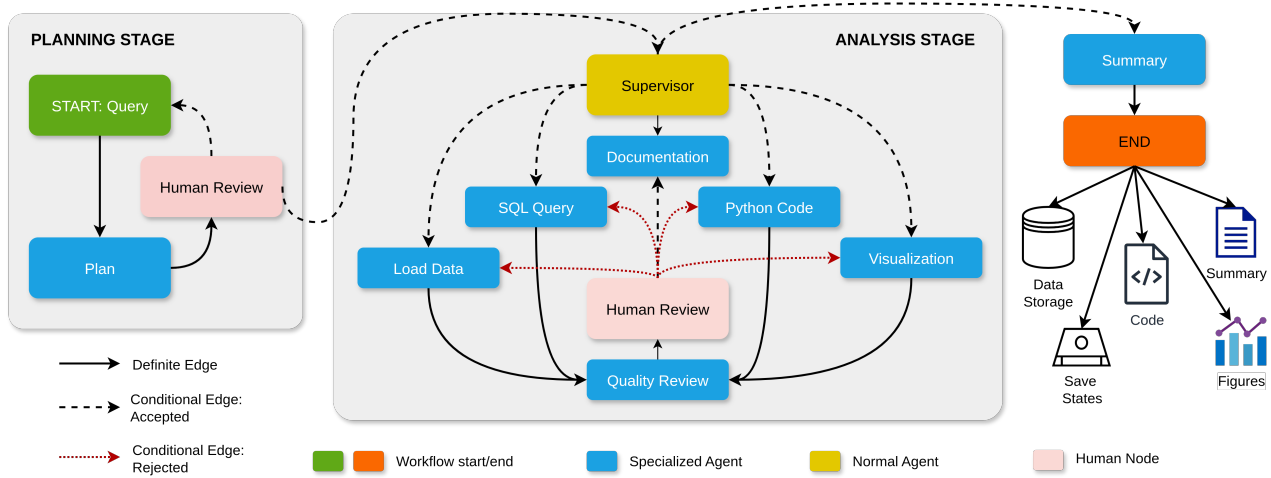

Figure 3: The InferA multi-agent architecture for scientific data analysis. The planning stage (left) involves iterative refinement of analysis plans through human feedback, while the analysis stage (middle) orchestrates seven specialized agents to execute the approved plan, transforming natural language queries into actionable insights and visualizations. The system produces provenance tracking output (right) including intermediate data, code, summary/documentation, and visualizations.

agent delegates tasks to seven specialized agents to complete the analytical process.

**Planning stage:** The planning stage implements a multi-turn dialogue module between the user and a dedicated planning agent. This agent employs chain-of-thought prompting to comprehend and extract the user's intent from their initial request. With comprehensive knowledge of all available agents' capabilities, the planning agent generates and refines a step-by-step analytical plan, decomposing complex queries into manageable components. This approach enables users to review, understand, and modify the plan as needed. The resulting plan serves as a road map for both the user and the downstream agents, ensuring a clear and executable strategy for addressing the task at hand.

**Analysis stage:** Following plan approval, the analysis stage begins under the direction of a supervisor agent, which orchestrates step-by-step task execution according to the established plan, while monitoring overall progress and performance. The supervisor interprets each step in the plan and delegates appropriate tasks to specialized agents.

The workflow starts with the data-loading agent, which is solely responsible for understanding the hierarchical structure of ensemble data such as HACC simulation runs. The data-loading agent assesses the entire ensemble context, including descriptions of each particle/property file, and determines which files and columns are necessary to load for all downstream tasks. This filtering reduces the required data from multiple terabytes to a few gigabytes at most. Selected data is written to a DuckDB database [23], avoiding in-memory storage, further minimizing storage-system overhead for provenance tracking, and facilitating access for downstream agents and user verification of intermediate results.

Once the database is created, an SQL programming agent performs additional filtering through generated SQL queries, evaluating whether all loaded columns and rows are necessary for immediate computation. For example, tasks evaluating a single target particle would require loading only the one corresponding row. The combined objective of the SQL programming agent, along with the data-loading agent, is to minimize storage and memory requirements for complex calculations.

A Python programming agent and visualization agent then work collaboratively to generate code for data analytics and visualization. Both code-generating agents implement basic functionalities using libraries like Numpy, Scipy or Matplotlib, though they have limited capabilities with domain-specific algorithms. To address this limitation, both agents incorporate multi-tool functionality and automated tool selection; custom algorithmic functions operating on pandas dataframes can be added to the system, and the agents will be able to apply these custom functions when appropriate. In our HACC dataset workflow, custom tooling enables halo tracking across time steps and facilitates ParaView [2] time-series visualization generation, both of which are domain-specific capabilities that would be too specialized and complex for an agent to develop.

Finally, a documentation agent maintains comprehensive records of operations, including AI-generated code and the successes and limitations encountered by each agent throughout the workflow.

### 3.1 Data Context with RAG

A critical component of our data loading process is RAG-enabled metadata extraction that identifies relevant column names from datasets. Scientific datasets often require domain expertise to interpret metadata, such as file contents and column labels. The HACC dataset exemplifies this challenge: column labels such as "sod_halo_MGas500c" are ambiguous labels representing complex concepts (in this case, "mass enclosed density 500 times the critical density in a spherical overdensity halo"). We address this through context-aware preprocessing that creates two dictionaries: one describing the ensemble file structure, and another mapping column labels to context-rich natural language descriptions. These dictionaries can be LLM-generated and expert-refined, enabling both easy exploration of new datasets with InferA and comprehensive context incorporation. The dictionaries serve as the knowledge base from which our system retrieves appropriate column names for downstream tasks.

| Semantic Complexity \ Analysis | Easy | Medium | Hard |
|---|---|---|---|
| **Easy** | Across all the simulations, what is the average size (fof_halo_count) of halos at each time step? | Please find the largest 100 galaxies and 100 halos at timestep 498 in simulation 0. I would like to plot all of them in Paraview and also see how well aligned those galaxies and halos are to each other. | Can you plot the change in mass of the largest friends-of-friends halos for all timesteps in all simulations? Provide me two plots using both fof_halo_count and fof_halo_mass as metrics for mass. |
| **Medium** | n/a | I would like to find the most unique halos in simulation 0 at timestep 498. Using velocity, mass, and kinetic energy of the halos, generate an 'interestingness' score and plot the top 1000 halos as a UMAP plot, highlighting the top 20 halos in simulation 0 that are the most interesting. | How does the slope and normalization of the gas-mass fraction–mass relation (sod_halo_MGas500c/sod_halo_M500c) evolve from the earliest timestep to the latest timestep in simulation 0? |
| **Hard** | n/a | First find the two largest halos by their halo count in timestep 624 of simulation 0. Then find the top 10 galaxies associated to those two halos (related by fof_halo_tag). What are the differences in characteristics of the two groups of galaxies? For example, differences in gas-mass, mass, or kinetic energy? | At timestep 624, how does the slope and intrinsic scatter of the stellar-to-halo mass (SMHM) relation vary as a function of seed mass? Which seed mass values produce the tightest SMHM correlation, and is there a threshold seed mass that maximizes stellar-mass assembly efficiency? |

Table 1: Difficulty matrix showing representative examples across analytical difficulty and semantic complexity. Questions are categorized as easy, medium, or hard. Analytical difficulty refers to the number of steps required to complete a query, while semantic complexity captures the degree of domain-specific knowledge needed. No representative questions were identified for Easy-Medium and Easy-Hard combinations.

Our retrieval approach avoids conventional size-based chunking, which typically divide text into fixed-length segments regardless of content boundaries. When applied to our dictionaries, conventional size-based chunking would merge unrelated column descriptions, significantly weakening similarity searches. Instead, we segment each column label into individual documents of at most 80 tokens. This fine-grained chunking strategy improves retrieval precision by returning highly relevant column names while avoiding dilution with unrelated context. To compensate for the smaller document size, our retriever employs maximum marginal relevance [7] to select the top 20 documents for several prompts: the original user query, the specific task assigned by the planning agent, the complete plan, and an "[IMPORTANT]" prompt that highlights columns tagged as important, retrieving up to 80 total documents.

## 3.2 Code Error Detection and Correction

While LLMs excel at code generation, they lack the native capability to execute and verify code, a safety feature that becomes limiting in analytical workflows. Our system implements an iterative loop for code testing, error checking, and quality assessment within a sandboxed environment isolated from the ground truth data, which guarantees that the original data will never be modified by the agent. The sandboxed environment is implemented as an asynchronous server gateway interface (ASGI) server using Uvicorn [8] from FastAPI [27], where the system transmits code and a temporary data copy to the server. The server executes the code, performs error detection, and returns either a complete error-free pandas dataframe or detailed error messages. Following execution, a quality assurance agent evaluates whether the output satisfactorily completes the delegated task.

## 3.3 InferA Evaluation Procedure

To evaluate InferA on the HACC dataset, we categorized 20 questions into 3 difficulty levels based on two metrics: **1) analysis complexity**, measured by the average number of steps in the analysis plans, and **2) semantic complexity**, assessed by the alignment between query terms and metadata descriptions.

For **analysis complexity**, questions were classified as easy ($< 4.5$ analysis steps), medium ($4.5 - 5.5$ analysis steps), and hard ($> 5.5$ analysis steps). Analysis steps specifically refer to operations performed during the data analysis phase, excluding planning, quality assurance, documentation, or summarization. Difficulty thresholds were established based on the typical analysis pipeline, which requires one data loading step plus one or more SQL, Python, and visualization operations. Thus, easier tasks generally involve four steps (one of each core analysis component), medium-difficulty require an additional computation or visualization step, and harder questions demand two or more additional operations.

For **semantic complexity**, "easy" questions contain terms directly defined in the metadata, requiring simple column extraction; "medium" questions use normalized wording not directly matching column names (e.g. "slope" or "normalization"); and "hard" questions include domain-specific terminology (e.g. "intrinsic scatter" or "velocity dispersion") absent from the metadata or that require contextual inference to determine which columns/characteristics are most relevant to the query. These hard questions demand deeper domain knowledge to disambiguate column selection and translate user intent into appropriate analysis without explicit guidance.

The example in Table 1 (bottom right) showcases a question classified as having both hard analysis difficulty and hard semantic complexity. The question is categorized as having hard analysis difficulty because across most runs, the planning agent decomposes it into eight distinct tasks: (1) load data, (2) filter for relevant columns (stellar mass, halo mass), (3) perform additional data cleaning in Python, (4) compute stellar-to-halo-mass (SMHM) relation, (5) create scatter plot of stellar mass vs. halo mass, (6) calculate intrinsic scatter of SMHM relation, (7) plot the intrinsic scatter for each seed mass, highlighting differences, and (8) find which seed mass produces the tightest SMHM correlation. With execution requiring an average of 7.7 steps across 10 runs (well above our 5.5 threshold), this question meets our criteria for high analysis difficulty. From a semantic perspective, this question is categorized as hard because it contains ambiguous terminology, such as "tightest correlation" and "intrinsic scatter", and relies on domain-specific concepts like "stellar-mass assembly efficiency" that aren't explicitly defined in the metadata. We can compare that to the question with hard analysis/medium semantic complexity (middle left) which asks for slope and normalization, concepts that are well-defined and not domain-specific.

We tested each question **10 times** without human feedback, either by skipping human feedback or instructing the LLM to "ignore missing requirements and continue." Skipping human feedback provides a lower bound for system performance, as the intended use case involves continuous human feedback at key steps. We established objective metrics to evaluate our multi-agent system across 10 runs per query, measuring operational efficiency and task fulfillment averaged across those 10 runs:

(1) **Data analysis success**—percentage of runs where the system extracted and computed data that directly addresses the specified task while maintaining topical relevance, supported by correct reasoning (this is to offset variability in interpretation) and produces valid output data.
(2) **Visualization success**—percentage of runs where the system produced on-topic visualizations with valid code that would generate valid visualizations if given accurate input data. This requires that the chosen form of visualization is reasonable for the data (e.g. 3D visualization for spatial analysis, line graph for time step analysis).
(3) **Reliability**—percentage of runs completing without failure.
(4) **Task Completeness**—average percentage of planned tasks successfully completed
(5) **Token usage**—average token consumption at termination.
(6) **Time taken**—average runtime to termination.

A fundamental challenge in assessing automated data analysis systems is that queries can have multiple valid interpretations, which complicates objective evaluation. We address this by measuring data analysis and visualization success based on alignment with explicit tasks assigned to system agents, rather than attempting to judge if results match implicit user intentions, which ensures consistent evaluation across diverse queries.

## 4 RESULTS AND DISCUSSION

We tested using OpenAI's GPT-4o with text-embedding-3-small for embeddings. GPT-4o was selected for its convenient API access, rapid response times, and widespread availability, despite its associated financial costs. While models like Anthropic's Claude may offer comparable or better performance in certain tasks, GPT-4o provided a good balance of accessibility and capability. Our testing confirmed that GPT-4o significantly outperforms locally-hosted security-compliant models available through Ollama, making the the former a necessary choice. All agents used custom-built prompts and routing, implemented using the LangChain library for basic agent tooling functionality and prompt chaining, and the LangGraph library for routing and state-based workflow management.

Following our evaluation on 20 questions across varying difficulty levels, tested on four HACC simulation runs (625 time-evolution snapshots each, 350 GB per run, totaling 1.4 TB) at Los Alamos National Laboratory, we observed the performance metrics shown in Table 2 and identified key patterns in how LLMs approach complex scientific reasoning tasks. Distribution of question difficulty is shown in Table 2 with question count indicating the number of questions out of 20 in that category. Additionally, for scalability purposes, we demonstrate one example with 32 simulation runs totaling 11.2 TB in Fig. 4 at Argonne National Laboratory.

### 4.1 Evaluations

*4.1.1 Task Completion Evaluation.* Task completion, defined as successfully passing all steps in the designated plan, was achieved in 85% of all runs, with performance correlating to both analysis difficulty (95%, 83%, 80% for easy, medium, and hard, respectively) and semantic complexity (91%, 92%, 74% for easy, medium, and hard, respectively). Even when runs did not complete, they made partial progress, with failed runs completing an average of 53% of their planned tasks. Across all runs (both successful and failed), our system completed 93% of all planned tasks overall.

These high completion rates highlight the effectiveness of our iterative quality assurance loop, which ensures valid code generation and execution even for complex analytical tasks. The quality assurance agent requests step revisions an average of 3.02 times per run, serving as a critical mechanism for maintaining code validity. The system's ability to maintain high reliability across increasing difficulty levels demonstrates how our multi-agent architecture successfully decomposes complex problems into manageable components, enabling the system to tackle challenges that would overwhelm simpler approaches while efficiently handling necessary revisions through targeted feedback.

Despite the system's overall effectiveness, we identified specific failure patterns in incomplete runs. The most common mechanism of failure involved using non-existent or slightly incorrect column names in the Python and visualization agents' code. While the quality assurance loop typically resolves these through error-guided iterations, multiple simultaneous errors in generated code would occasionally cause the system to reach the maximum threshold of five revision attempts without resolving all issues. Importantly, these syntactic errors are quickly identified and easily resolved through human feedback, suggesting that a modest amount of human oversight can substantially enhance the system's already high completion rate.

*4.1.2 Quality Assessment Evaluation.* Our quality assessment revealed that 76% of all runs resulted in valid and satisfactory data analysis outcomes, while 72% of all runs produced satisfactory visualization outputs. Considering that 15% of unsatisfactory runs were due to early termination, we found that the system is relatively successful in producing satisfactory data outcomes (85%) and visualization outcomes (80%) when runs are completed. However, a minor portion of runs that completed translated to unsatisfactory analysis choices.

|  | Difficulty (count) | % Satisfactory Data | % Satisfactory Visual | % of Runs Completed | % Complete | Token Usage | Storage Overhead (GB) | Time (s) | Redo Iterations |
|---|---|---|---|---|---|---|---|---|---|
| **Analysis Difficulty** | Easy (6) | 87% | 78% | 95% | 98% | 87907 | 1.51 | 279 | 1.48 |
|  | Medium (6) | 67% | 67% | 83% | 92% | 119266 | 0.09 | 211 | 3.80 |
|  | Hard (8) | 76% | 71% | 80% | 89% | 139464 | 1.86 | 543 | 3.59 |
| **Semantic Complexity** | Easy (8) | 81% | 78% | 91% | 96% | 94184 | 1.37 | 270 | 1.43 |
|  | Medium (5) | 77% | 78% | 92% | 97% | 124880 | 0.99 | 301 | 1.77 |
|  | Hard (7) | 70% | 61% | 74% | 85% | 140125 | 1.22 | 517 | 5.74 |

| # Simulation | Timestep |  |  |  |  |  |  |  |  |
|---|---|---|---|---|---|---|---|---|---|
| Single | Single (7) | 69% | 77% | 86% | 93% | 108428 | 0.04 | 174 | 2.90 |
|  | Multi (5) | 88% | 62% | 94% | 98% | 101269 | 3.18 | 739 | 1.76 |
| Multi | Single (5) | 70% | 64% | 74% | 86% | 145591 | 0.21 | 263 | 5.54 |
|  | Multi (3) | 86% | 90% | 90% | 98% | 121817 | 2.41 | 353 | 1.21 |

| **Total** | (20) | 76% | 72% | 85% | 93% | 117938 | 1.22 | 364 | 3.02 |

| **Successful runs** |  | 85% | 80% | 100% | 100% | 109175 | 1.26 | 336 | 1.49 |
| **Unsuccessful runs** |  | 17% | 17% | 0% | 53% | 166596 | 0.65 | 485 | 12.09 |

Table 2: Performance evaluation of the InferA multi-agent system across 200 runs (20 questions, 10 runs each). The table presents metrics categorized by task difficulty, multi-simulation/multi-timestep requirements, and success status, showing completion rates, quality assessments and resource utilization patterns across the three categories.

Issues with unsatisfactory outputs mostly stemmed from the LLM selecting inappropriate analytical techniques or visualization formats rather than from code execution failures. A frequent issue is asking the LLM to track the evolution of characteristics rather than particles (e.g. scalar value, mass) and the LLM incorrectly uses the particle coordinate tracking tool, resulting in valid but unsatisfactory output. This finding suggests that while LLMs can translate basic intentions into executable code, they still struggle with higher-level scientific reasoning needed to interpret complex intentions and apply appropriate analytical approaches. This underscores the importance of human-in-the-loop validation to ensure that the generated approach aligns with scientific question being investigated.

*4.1.3 Computational Efficiency and Resource Utilization.* In comparison to the test dataset, which is 1.4 TB in size, the storage overhead for provenance tracking ranged between 8 MB and 4.9 GB, with the upper range only reached when answering questions that require data from multiple time steps with many variables. Questions querying only one timestep required substantially less storage overhead (8 MB to 428 MB). Even the maximum average storage overhead represents less than 0.35% of the full dataset. Note also that the majority of the data is stored in DuckDB databases, which perform data operations on disk rather than in memory, making the actual memory requirements an almost negligible fraction of the full dataset. For context, the example workflow with 32 simulation runs (11.2 TB total) produced a database of 18 GB and 5 pandas dataframes of 1.4 MB each.

This dramatic reduction in storage requirements represents one of InferA's most significant contributions. By intelligently subsetting and caching only the relevant portions of data, the system makes large-scale cosmological analyses accessible on standard computational hardware. Traditional approaches would require either substantial HPC resources to wrangle large data or significant time investments from researchers for manual data reduction.

The average execution time for each query ranged between 96 seconds and 1412 seconds, with analyses that loaded the most data requiring the most execution time. Notably, despite runs taking minutes to complete, only a small fraction of this time is related to LLM invocation with no invocation taking more than 5 seconds with GPT-4o; the majority of time is instead consumed by Python code execution on the data itself, an unavoidable cost that could only be reduced by intelligent, manual coding. The runtime metric is only an indicator of the average wait time while exploring with InferA rather than an evaluation of the system itself.

*4.1.4 Token Usage.* Token usage averaged 117,938 across all runs and ranged between 65,000 and 178,000 tokens per query. This average is notably skewed upward by failed runs, which consumed significantly more computational resources (averaging 166,596 tokens) compared to successful runs (averaging 109,175 tokens) despite there being many fewer failed runs. Although the costs are relatively high for a single query, the cost is favorable considering the complexity of the analyses performed and the human time saved. We observed that token usage increased with both the complexity of the analysis and the number of iterations required to reach a satisfactory result, suggesting that a human-in-the-loop workflow would reduce token cost and that future optimizations should focus on improving first-pass analysis quality.

There are several points at which token usage can be lowered. For simpler queries where sub-tasks are straightforward, lowering the message history passed to the supervisor agent drastically reduces token usage. Similarly, the documentation agent functions as a workflow summarization tool that captures processing steps for easy review without contributing to analysis or provenance tracking. While useful for understanding the system's operations, this documentation feature is not strictly necessary for core analysis.

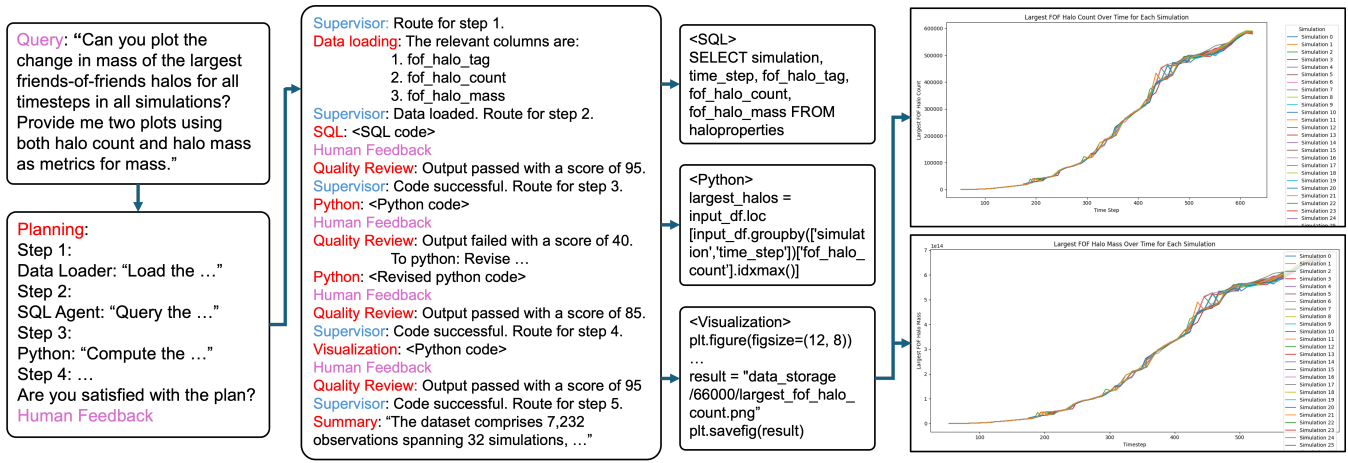

**Figure 4: Example workflow with query asking to plot the halo count and halo mass for 32 simulations over all timesteps.**

### 4.2 Key Features

*4.2.1 Provenance Tracking and Stateful Architecture.* Comprehensive provenance tracking is essential for scientific reproducibility in data-analysis workflows. By systematically recording all intermediate comma-separated value (CSV) files, executed code, and generated outputs in sequential order, the system creates a complete audit trail of the analytical process. This detailed documentation makes experiment replication straightforward, allowing researchers to recreate and verify analytical pathways with minimal effort. The provenance records serve as transparent evidence of the LLM's chain-of-thought reasoning process, enabling verification of results while supporting scientific accountability.

Complementing this, InferA's stateful architecture provides a crucial feature for experimental workflows: by capturing and preserving the exact computational state from each analysis agent, the system enables efficient workflow branching and exploration. This statefulness allows analysts to load from specific checkpoints and alter follow-up steps, significantly reducing computational overhead when testing alternative approaches. Rather than rerunning entire workflows, researchers can branch from established processing stages to explore different analytical paths, facilitating comprehensive exploration of datasets while conserving computational resources and time.

*4.2.2 Human-in-the-loop.* While our evaluation procedure focused on fully automated conditions, InferA was designed with human collaboration as a central feature, allowing the user to adjust output on the fly. Although not evaluated in our test suite, there is discernible improvement in the efficiency of the system to reach accurate results when human feedback is provided. Of the runs that executed valid code, some selected suboptimal analytical approaches with 9% producing unsatisfactory data and 13% producing unsatisfactory visualization. 15% of runs failed due to errors such as incorrectly-named data tables, columns, or non-existent functions—issues readily addressable through targeted human intervention. For example, when agents incorrectly simplify variable names (e.g., using $center\_x$ instead of $fof\_halo\_center\_x$), directly providing the correct name resolves the issue, avoiding multiple correction attempts. As a result, we believe the numbers in our evaluation metrics to be a lower bound for actual reliability and accuracy.

*4.2.3 Sandboxed Execution Environment.* InferA addresses a critical gap in current data analysis assistants by implementing a secure execution environment that separates code generation from execution. Many current systems do not provide plug-and-play sandbox execution, often inseparably packaging code generation with execution. Our architecture executes all code on temporary data copies within an isolated server environment with error-checking mechanisms, providing robust protection of source data integrity while enabling unrestricted analytical exploration.

*4.2.4 Error Detection.* Our implementation of the quality assurance agent revealed that binary correctness assessments of code (correct/incorrect) frequently lead to false negatives, even if the code perfectly executes the task. To avoid this, we implemented a nuanced scoring approach where the quality assurance agent assigns a score on a scale of 1-100 without rigid criteria. This approach, with a threshold of 50 for correct/incorrect determination, proved significantly more effective at lowering false negatives while still catching subtle issues in code that would otherwise appear superficially correct.

*4.2.5 Agent interaction and communication.* The division of tasks among specialized agents enables optimized inter-agent communication patterns throughout the system. With the exception of the supervisor agent, each agent operates with limited context awareness, receiving only its delegated task without knowledge of upstream processes. This approach maintains functional efficiency while significantly reducing token costs, and does not compromise agents' ability to effectively complete their assigned tasks.

### 4.3 Case studies

We demonstrate InferA's capabilities through an example analysis requiring multiple simulations and multiple time evolution steps. The query requests the creation of two plots from all 32 simulations, visualizing the halo count and halo mass of the largest halo from all time steps. Through five analysis steps—loading data, creating a filtered data base, organizing data by timestep, and two separate visualizations for each plot—InferA created two matplotlib figures shown in Fig. 4. The original 32 simulations totaled 11.2 TB; in comparison, the storage overhead consisted of a database at 18 GB and CSVs loaded in-memory that averaged 1.4 MB. The total run time was 5403 seconds and the entire process used a total of 126,568 tokens. This example represents a significant success in manipulating multi-terabyte data sets in an automated workflow with no concern for memory overload.

In another example (Fig. 5), InferA used custom tools to generate a 3D visualization, a task the base LLM frequently performs incorrectly. The query requested visualization of a target dark matter halo and all surrounding halos within a 20 megaparsec radius. The target halo was successfully highlighted in red using Paraview, demonstrating InferA's ability to integrate specialized tools.

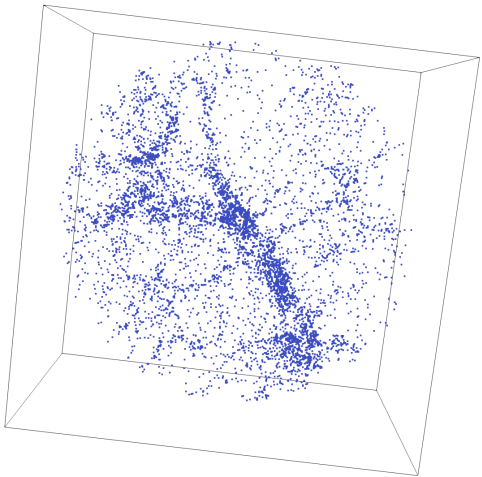

**Figure 5: Example Paraview visualization of a target dark matter halo and all halos within 20 Mpc of it.**

## 4.4 Comparative Performance Assessment

Our exploratory comparison with baseline approaches, including direct LLM chat models and PandasAI, revealed significant limitations in handling complex cosmological ensemble data. Standard chat models quickly exceeded context windows even with toy data samples: a 20 × 5 dataframe already resulted in hallucinated values and relationships. Similarly, PandasAI proved incompatible with ensemble analysis requirements, failing to extract meaningful column semantics and unable to process the necessary data volumes.

*4.4.1 Multi-Agent vs. Alternative Architectures.* The multi-agent approach demonstrated clear advantages over both single-system implementations and static linear workflows. By decomposing complex tasks into specialized functions, InferA successfully navigated analytical challenges that overwhelm simpler architectures. The dynamic task allocation enabled by the supervisor agent also allowed the system to flexibly adapt to queries' differing analytical requirements. We found that agent-generated Python code consistently outperformed specialized pandas code in both execution reliability and analytical complexity. By instructing the agent to generate Python code compatible with the pandas library, the agent delivered solutions with superior flexibility and extended capabilities compared to the pandas-specific implementations.

## 4.5 Analytical Variability and Interpretation

Our evaluation revealed substantial variability in how the system interpreted and approached questions, especially those containing inherent ambiguities. For example,

> *Can you make an inference on the direction of the FSN and VEL parameters in order to increase the halo count of the 100 largest halos in timestep 624? Also plot a summary of the differences in halo characteristics between the two simulations.*

This question contains significant ambiguities that, without clarification, introduce many possible analytical pathways. Specifically, phrases like "direction of the parameters" and "halo characteristics" lack precise definition, providing no straightforward analytical direction. Consequently, the system explored multiple valid analytical strategies across different runs for the same question:

- Calculating average mass differences among the 100 largest halos and averaging parameters;
- Determining linear correlations between simulation parameters and halo masses via linear regression;
- Performing one-to-one comparisons between corresponding halos; and
- Generating correlation matrices across characteristic variables for each simulation.

This analytical diversity typically emerges when questions are ambiguous. In contrast, when presented the following precise, unambiguous query that targets only one particle and one characteristic:

> *Can you find me the top 20 largest friends-of-friends halos from timestep 498 in simulation 0?*

InferA successfully produced identical data outputs and very similar visualizations and analytical approaches across all 10 runs.

## 5 CONCLUSION AND FUTURE WORK

In this paper we introduced InferA, a multi-agent smart assistant for large cosmology datasets from the HACC simulation. By leveraging LLMs and a domain-specific RAG interface, InferA translates natural-language queries into structured data operations, mining relevant information directly from disk. To demonstrate scalability, we ran InferA on an 11.2 TB ensemble of 32 simulation runs at ANL; a dataset too large to fit into an LLM's context window or memory-constrained tools such as PandasAI. The system efficiently reduced storage requirements to less than 0.35% of the original size and minimized in-memory loading, allowing complex analyses on standard computational hardware. Full provenance tracking ensures that every analytical decision can be reviewed, verified, and reproduced, preserving scientific rigor. Moreover, InferA's modular design allows straightforward adaptation to other domains by implementing data-specific data loaders and metadata dictionaries. Its multi-agent architecture decomposes complex analyses into manageable steps, while sandboxed execution ensures data integrity, enabling data manipulation without risk of corruption.

As future work, we would like to integrate a web agent that can review online data sources about cosmology and then better address questions by combining data from the HACC simulation as well as information from publications, and investigate parallelized workflow execution to reduce execution runtime.

## ACKNOWLEDGMENTS

Work at Los Alamos National Laboratory was supported by the U.S. Department of Energy, Office of Science, Office of Advanced Scientific Computing Research and Office of DE-SC0021399, Scientific Discovery through Advanced Computing (SciDAC) program. Work at Argonne National Laboratory was supported under the U.S. Department of Energy contract DE AC02-06CH11357. The authors would like to thank Kei Davis for his diligent copy editing of this paper, Michael K. Lang, John Patchett, Tom Berg, Mark Myshatyn, and Becky Rutherford for providing us access to AI resources at LANL. The authors thank Nicholas Frontiere, Michael Buehlmann, JD Emberson, and the HACC team for generating and providing the ensemble of cosmological simulations. We also thank Salman Habib, Katrin Heitmann and Azton Wells for useful discussions related to this work.